\documentclass[twocolumn,showpacs,aps,prd,amsmath,amssymb,nofootinbib,nobibnotes,floatfix]{revtex4-1}

\usepackage{hyperref}
\usepackage{amsfonts}
\usepackage{amsmath}
\usepackage{mathrsfs}
\usepackage{epsfig}
\usepackage{graphicx,epstopdf}               
\usepackage{url}
\usepackage{hyperref}
\usepackage{float}
\usepackage{color}
\usepackage[utf8]{inputenc} 
\usepackage{enumerate}
\newcommand{\Alfven}{Alfv{\'e}n }
\newcommand \bh {_{\rm BH}}
\usepackage{graphicx}
\usepackage{epsf}

\newcommand{\ssec}[1]{\section{#1}}
\usepackage{comment}

\newcommand{\nc}{\newcommand}

\nc{\beq}{\begin{equation}}
\nc{\eeq}{\end{equation}}
\nc{\beqa}{\begin{eqnarray}}
\nc{\eeqa}{\end{eqnarray}}

\usepackage{slashed}

\newcommand{\lsim}{\!\mathrel{\hbox{\rlap{\lower.55ex \hbox{$\sim$}} \kern-.34em \raise.4ex \hbox{$<$}}}}
\newcommand{\gsim}{\!\mathrel{\hbox{\rlap{\lower.55ex \hbox{$\sim$}} \kern-.34em \raise.4ex \hbox{$>$}}}}

\def\be{\begin{equation}}
\def\ee{\end{equation}}

\newcommand\affspc{\vspace{4pt}}

\usepackage{footmisc}
\usepackage{setspace}
\begin{document}

\title{Fate of a neutron star with an endoparasitic black hole and implications for dark matter}

\author{William E.\ East}
\affiliation{Perimeter Institute for Theoretical Physics, Waterloo, Ontario N2L 2Y5, Canada \affspc}
\author{Luis Lehner}
\affiliation{Perimeter Institute for Theoretical Physics, Waterloo, Ontario N2L 2Y5, Canada \affspc}

\begin{abstract}
    We study the dynamics and observational signatures of a neutron star being
    consumed by a much less massive black hole residing inside the star. This
    phenomenon could arise in a variety of scenarios, including after the 
    capture of a primordial black hole, or in some models of asymmetric dark
    matter where the dark matter particles collect at the center of a neutron
    star and eventually collapse to form a black hole.  However, the details of
    how the neutron star implodes are not well known, which is crucial to determining the
    observational implications of such events.
    We utilize fully general relativistic simulations
    to follow the evolution of such a black hole as it grows by several orders
    of magnitude, and ultimately consumes the neutron star. We consider a range
    of spin values for the neutron star, from nonrotating stars to those with
    millisecond periods, as well as different equations of state. We find
    that as the black hole grows, it obtains a non-negligible spin and induces differential
    rotation in the core of the neutron star. In contrast to previous studies, we find
    that the amount of dynamical ejecta is very small, even for rapidly rotating
    stars, dampening the prospects for producing a kilonova-type electromagnetic signal
    from such events.
    We comment on other possible electromagnetic and gravitational signals.
\end{abstract}

\maketitle

\ssec{Introduction}%
Recently, there has been much interest in scenarios where neutron stars (NSs)
capture dark matter---either particles or primordial black holes (BHs)---giving
rise to a tiny BH which consumes the NS from the inside through accretion.
This could give rise to a number of observable consequences, allowing
NSs to play the role of dark matter detectors, while addressing several astrophysical
mysteries, as we review below.

This phenomenon could arise due to bosonic or fermionic asymmetric dark
matter~\cite{Goldman:1989nd,deLavallaz:2010wp,Kouvaris:2010jy,McDermott:2011jp,Bramante:2014zca,Bramante:2015dfa}.
The dark particles are captured by the NS due to neutron
scattering, and if their annihilation and decay rates are sufficiently small,
they will eventually thermalize and collect to form a tiny ($\lesssim10^{-10}\
M_{\odot}$) BH at the NS center~\cite{Goldman:1989nd,Bramante:2017ulk}. Similarly, if primordial
BHs with masses $\sim10^{-14}$--$10^{-8}\ M_{\odot}$ make up a fraction of
the dark matter, they will be captured onto NSs through dynamical friction
and accretion~\cite{Capela:2013yf}.
In either case, the BH should grow through accretion until it consumes the NS.

The BH-induced implosion of NSs, which would preferentially
occur in regions of high dark matter density, has been invoked as a possible
explanation for the scarcity of observed pulsars in the Galactic center~\cite{Bramante:2014zca}.  
It has also been proposed that this could give rise to fast radio bursts (FRBs) due to
the rapid expulsion of the NS magnetosphere~\cite{Fuller:2014rza} 
(though this type of cataclysmic event cannot explain repeating FRBs).
Finally, in Refs.~\cite{Bramante:2016mzo,Fuller:2017uyd}, it was proposed that ejecta from the imploding
NS could explain the production of the majority of heavy elements through the r process.

There are several observational signatures that could be used to identify
such events. One would be the presence of $\sim 1\ M_{\odot}$ BHs 
in binary merger events detected by LIGO/Virgo and other gravitational
wave (GW) observatories (though NS implosions may not be the only conceivable explanation). 
Actually distinguishing BHs from NSs of the same mass with the merger GW signal
is challenging because tidal effects are significant only in the high frequency
part of the signal where current detectors are not very
sensitive~\cite{TheLIGOScientific:2017qsa}, and because if at least one 
of the constituents of the binary is a NS, there is a degeneracy in
the leading order tidal effect due to our ignorance about the equation of state
(EOS) describing NSs~\cite{Yang:2017gfb}. 
If detected, electromagnetic
counterparts to such GW signals could be used to provide strong evidence of the
presence of a NS, though ruling out a BH-NS binary is more
difficult~\cite{Yang:2017gfb,Hinderer:2018pei,Foucart:2019bxj}.

Another possible signature coming from the collapse of a NS to a BH is a
kilonova: an IR/optical/UV transient powered by the radioactive decay of
ejected, neutron-rich material.  However, the amount of ejecta depends
crucially on the dynamics of process, and on the distribution of angular
momentum as the BH grows, which is not well understood.  Most previous studies have left
this amount as an unknown, bounded above by measurements of the total abundance of
r-process elements.  An exception to this is Ref.~\cite{Fuller:2017uyd}, where
it was estimated that the collapse of a millisecond period
NS could produce $\sim0.1$ to $0.5\ M_{\odot}$ in ejecta, a range of values
in excess of those found in binary NS mergers~\cite{Siegel:2019mlp}. 

Several works have estimated the timescales and relevant physical processes
involved during the months to years-long period where the BH grows through
accretion inside the NS, but has negligible effect on the star as a whole.  (We
note that for the BH mass ranges mentioned above, the Hawking evaporation time is
much longer than the age of the Universe.) In Refs.~\cite{Kouvaris:2013kra,Fuller:2017uyd}, 
building from the related problem of a BH inside a solar-type star~\cite{Markovic:1994bu},
it was argued that the effects of nuclear viscosity and magnetic braking would be important.
Based on Newtonian estimates, these works contend that spherical (i.e. Bondi-like) accretion is  
maintained throughout the whole consumption process. Such behavior is governed by
sheer viscous braking for BH masses $\lesssim 10^{-3} M_{\odot}$, and magnetic braking for 
larger masses.

In this work, we study how the BH grows from a mass that is small, but
near becoming dynamical important---we begin
our analysis when the BH is roughly a hundredth of the NS mass---,to ultimately
consume the majority of the NS, a process which occurs over a few milliseconds.  
We explore several initial conditions based on uniformly rotating star solution---thus making contact
with previously mentioned estimates but studying its future evolution within general relativity.
We use solutions of the coupled Einstein-hydrodynamics equations, which allows us
to follow the relativistic dynamics as the BH begins to backreact on the rotating
NS, and make accurate estimates of potential observational signatures.  
The large disparity between the length scales respectively associated with the NS
and BH makes this a computationally challenging problem, which we overcome
through a combination of several techniques, including the use of adaptive mesh
refinement with flux corrections for the hydrodynamics, and by exploiting
spacetime symmetries.

We find that the BH develops and maintains a non-negligible spin, and induces
differential rotation in the core of the NS, and we give a simple heuristic
description for the accretion of angular momentum as the BH grows.
The rate of mass accretion, however, is largely insensitive to angular momentum.
We show that essentially no matter remains outside the BH
horizon at the end of the process, even for NSs rotating at speeds near breakup,
which means that these events are not promising sources for 
kilonovae or r-process heavy elements. 
We comment on possible GW and electromagnetic signatures, including bursts from 
the sudden release of the energy stored in the NS's magnetosphere.

\ssec{Methodology}%
In order to study the dynamics of a rotating NS being consumed by a BH from the inside,
we solve the Einstein equations coupled to hydrodynamics.  To construct initial
data, we begin with a uniformly rotating NS solution obtained using the RNS
code~\cite{Stergioulas:1994ea}.  We then alter this solution in a neighborhood
of the center of the star to match on to a small BH metric 
with a quasiequilibrium test fluid.
We use this as free data for solving the Einstein constraint equations as
described in Ref.~\cite{idsolve_paper}.  Though this solution is only an approximate
description of the system of interest, 
we have tested several initial values for
the BH mass $M_{\rm BH}$ in order to establish that we start with a
sufficiently small value that any initial transients are negligible.
More details 
are provided in Appendix~\ref{sec:ics}. 

The evolution is carried out using the methods (and code) described in
Ref.~\cite{code_paper}.  This includes the use of adaptive mesh refinement with
flux corrections~\cite{bc89}, which ensure that the conservative nature of the
hydrodynamic evolution scheme is not broken by the numerous levels of mesh
refinement with boundaries inside the NS which are required to resolve the BH. 
The hydrodynamic scheme conserves rest mass and, in axisymmetry, angular momentum.

We fix the NS to have a mass of $M_{\rm NS}=1.4\ M_{\odot}$ and consider three
different EOSs. We use the SLy and ENG EOS, which are ``softer" and give radii of
11.7 and 12 km, respectively, for a nonrotating NS. 
We also use the stiffer H4 EOS, which gives a radius of 14 km for a
nonrotating NS, and is at the edge of the allowed range consistent with the
binary NS merger GW170817~\cite{Abbott:2018exr}.  In particular, we use the
piecewise polytrope approximation of these EOSs described in
Ref.~\cite{Read:2008iy}.  As we describe below, the systems we consider develop
densities in the vicinity of the accreting BH that are larger than the maximum
density of isolated NSs (even ones of $\sim 2\ M_{\odot}$), 
and the softer EOSs, when extrapolated to these
values, would give superluminal sound speeds. To address this, at the value of
the rest-mass density when the sound speed becomes $c_s \approx 0.99$
($\rho_0=1.43\times10^{15}$ gm/cm$^3$ for ENG; $\rho_0=1.96\times10^{15}$
gm/cm$^3$ for SLy), we match onto an EOS with $c_s$ held constant at this
value.  (Unless otherwise specified, we use geometric units with $G=c=1$
throughout.) To this cold EOS, we also add a thermal component with $P_{\rm
hot}=0.5 \rho_0 \epsilon_{\rm hot}$, where $\epsilon_{\rm
hot}=\epsilon-\epsilon_{\rm cold}$ is the specific energy is excess of the
value prescribed by the cold EOS~\cite{Bauswein2010}. However, we find that in all the
cases we study, shock heating effects are not significant.

\begin{table}[t]
\caption{\label{rns_table} 
 }
 Properties of equilibrium uniformly rotating NSs used to construct ID.
    Left to right, the columns contain the EOS, dimensionless NS spin,
    rotational period, equatorial radius, and ratio of angular  
    velocity to that of a particle orbiting at the equator.
\centering
\begin{tabular}{ccccc}
\hline\hline
    EOS &
    $a_{\rm NS}$ &
    Period (ms) &
    $R_{\rm eq}$ (km) &
    $\Omega/ \Omega_K$
\\
\hline
    ENG & 0.00 & --- & 12.0 & 0.00 \\
    ENG & 0.10 & 5.3 & 12.0 & 0.12 \\
    ENG & 0.20 & 2.7 & 12.2 & 0.23 \\
    ENG & 0.40 & 1.5 & 12.9 & 0.47 \\
    ENG & 0.70 & 1.0 & 16.5 & 0.95 \\
    H4  & 0.00 & --- & 14.0 & 0.00 \\
    H4  & 0.10 & 6.6 & 14.0 & 0.12 \\
    H4  & 0.20 & 3.4 & 14.2 & 0.23 \\
    H4  & 0.40 & 1.8 & 15.1 & 0.47 \\
    H4  & 0.67 & 1.3 & 18.7 & 0.89 \\
    SLy & 0.00 & --- & 11.7 & 0.00 \\
    SLy & 0.20 & 2.6 & 11.9 & 0.24 \\
    SLy & 0.40 & 1.4 & 12.7 & 0.48 \\
\hline\hline 
\end{tabular}
\end{table}

For each EOS, we consider a range of values for the dimensionless NS spin
$a_{\rm NS}$, from nonspinning to millisecond period NSs near breakup, as
shown in Table~\ref{rns_table}.
For most cases, and unless otherwise stated, we choose the initial BH to have
mass $M_{\rm BH}=10^{-2}M_{\rm NS}$ and dimensionless spin $a_{\rm BH}=a_{\rm NS}$, and we assume
axisymmetry. Here and throughout we use $M_{\rm NS}$ to refer to the total mass
of the spacetime, which is equal to the NS mass before the BH obtains a non-negligible
mass.
For the case with the ENG EOS and $a_{\rm NS}=0.4$, we consider
several variations as a check of our initial conditions and numerical errors.
We study larger initial BHs with $M_{\rm BH}/M_{\rm NS}=0.02$ and 0.03. We
examine different initial BH spins with $a_{\rm BH}=0$, 0.2, and 0.4.  To
establish convergence and estimate numerical errors, we also adopt several
numerical resolutions (with $a_{\rm BH}=0.4$ and $M_{\rm BH}/M_{\rm
NS}=10^{-2}$).
Finally, we consider one case calculated without assuming any spatial symmetries,
starting from initial conditions from the axisymmetric simulations
at the time when $M_{\rm BH}/M_{\rm NS}=0.05$, in order to check for nonaxisymmetric
instabilities.
As detailed in Appendix~\ref{sec:conv}, in all cases we obtain similar results.

During the evolution, we track the BH apparent horizon and
monitor its area $A_{\rm BH}$ and angular momentum (from which the BH mass $M_{\rm BH}$
and dimensionless spin $a_{\rm BH}$ is calculated,
via the Christodoulou formula). We also calculate the flux of conserved
matter quantities into the horizon: the rest-mass accretion rate $\dot{M}_0$
and the angular momentum accretion rate $\dot{J}$.

We also determine the amount of unbound rest mass by integrating all
the fluid cells where the lower time component of the four velocity $u_t<-1$,
and the radial component of the velocity is outward, as is typically done 
in NS merger simulations.

\begin{figure}
\begin{center}
\includegraphics[width=\columnwidth,draft=false]{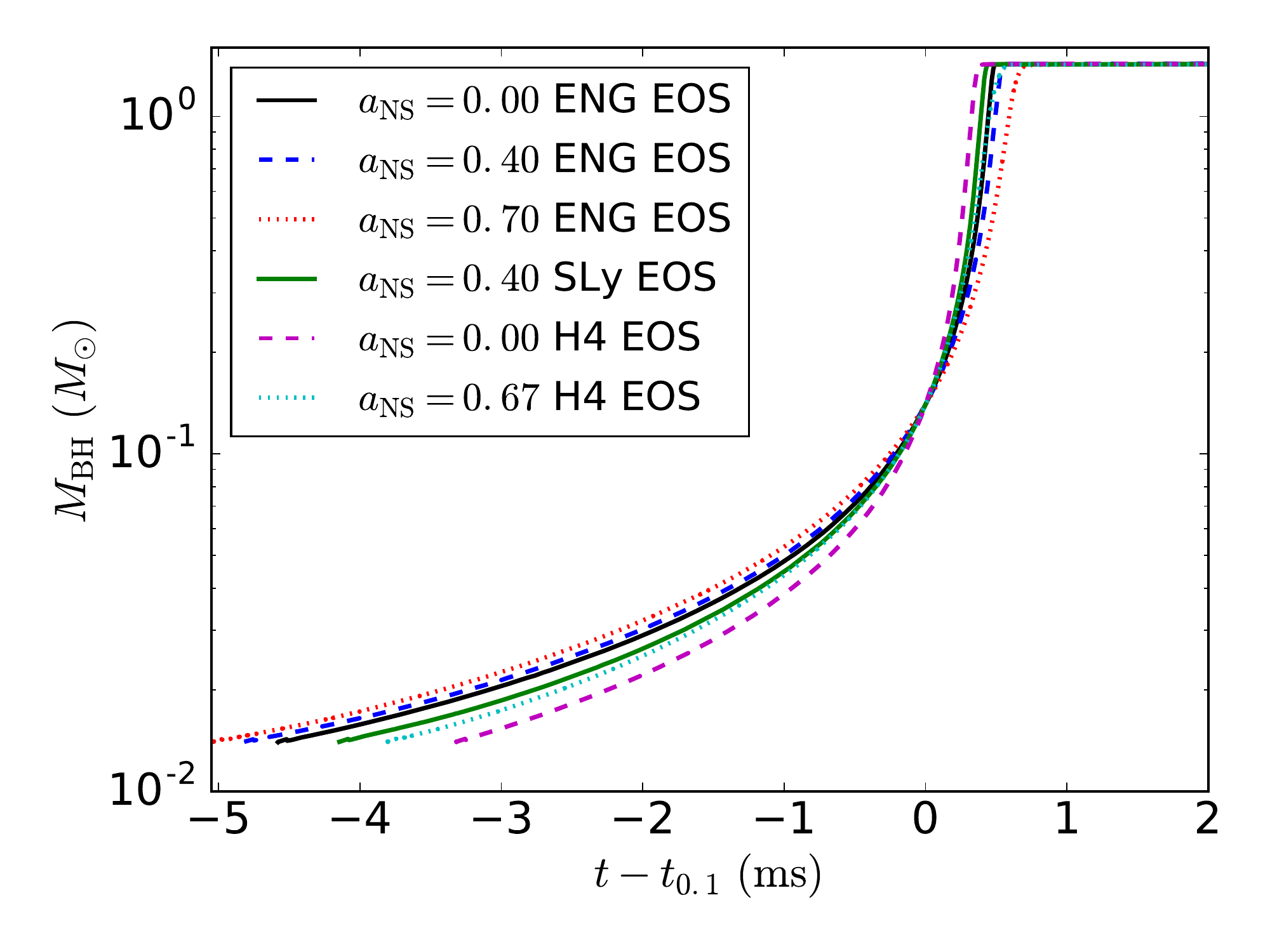}
\includegraphics[width=\columnwidth,draft=false]{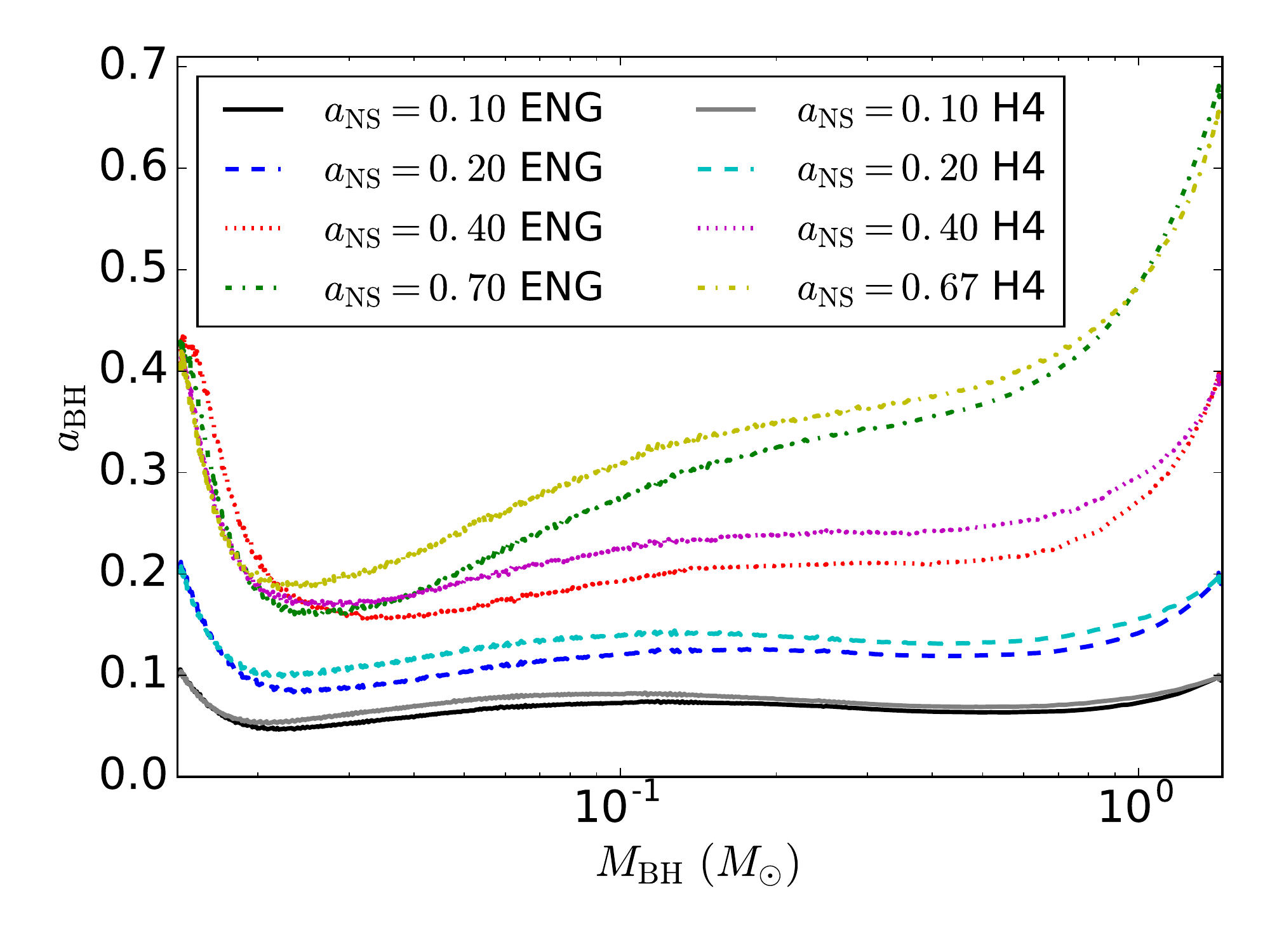}
\end{center}
\caption{
    Top: The mass of the BH as a function of time for cases with different 
    spins and EOSs for the NS.
    The curves have been aligned at the time where $M_{\rm BH}=0.1M_{\rm NS}$.
    Bottom:
    The dimensionless spin of the BH versus its mass.
\label{fig:mbh_comp}
}
\end{figure}

\ssec{Results}
Starting from one one-hundredth the mass of the NS, we find that the ``endoparasitic" BH efficiently grows to
consume essentially the entire NS in $\approx4$--6 ms. As shown in the
top panel of Fig.~\ref{fig:mbh_comp}, the accretion rate of the BH is mostly
determined by the NS EOS. The spin of the NS has only a small effect on the
growth rate of the BH, with higher spins giving slightly slower growth rates. 
The spin of the NS does, however, affect the BH spin. As evident in the bottom
panel of Fig.~\ref{fig:mbh_comp}, $a_{\rm BH}$ settles to a value of 
$\sim a_{\rm NS}/2$, and roughly maintains this value as the BH mass grows more than 
an order of magnitude. At the last stages of the NS being consumed, $a_{\rm BH}$
jumps up to match $a_{\rm NS}$.

\begin{figure}
\begin{center}
\includegraphics[width=\columnwidth,draft=false]{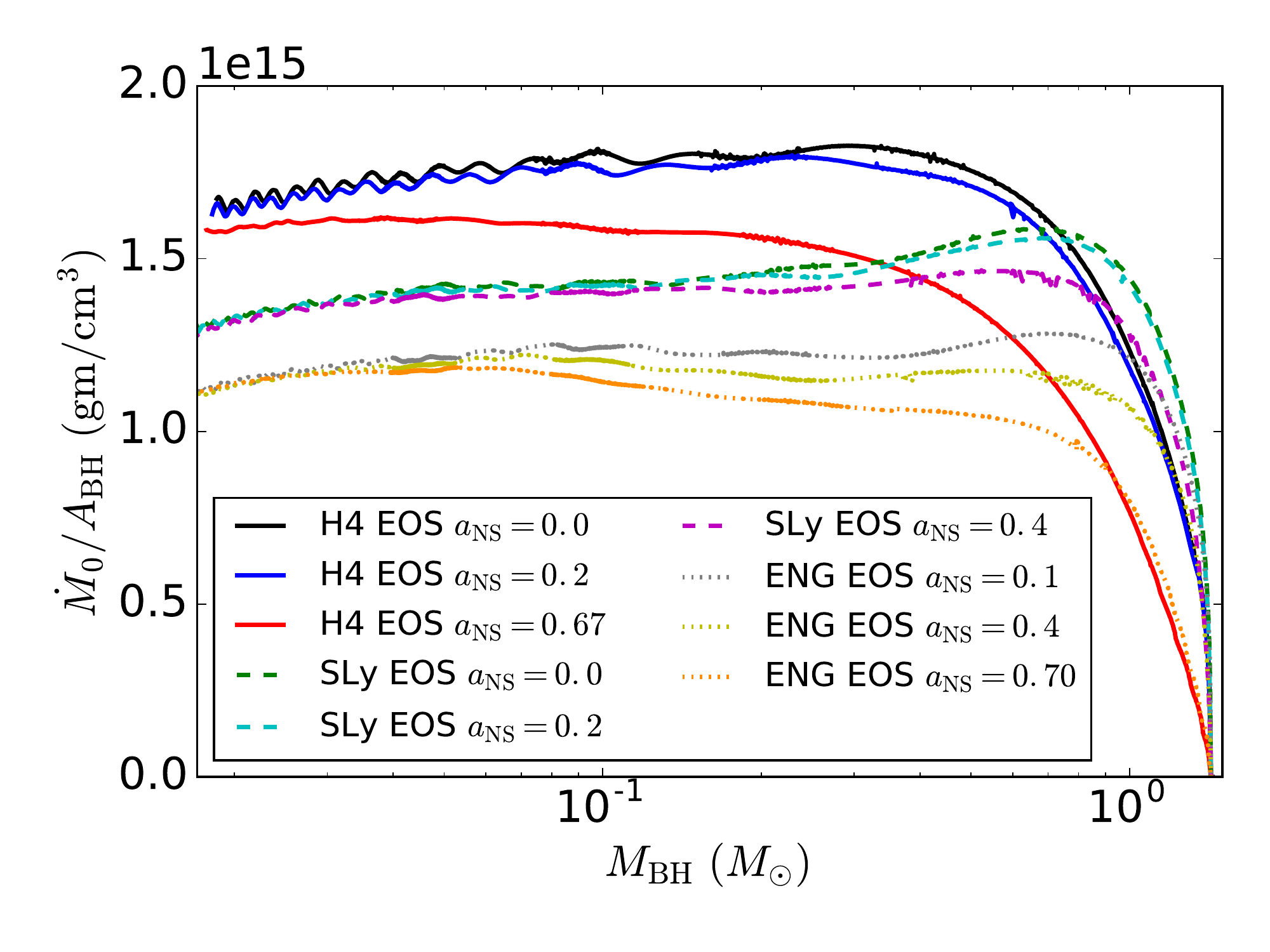}
\includegraphics[width=\columnwidth,draft=false]{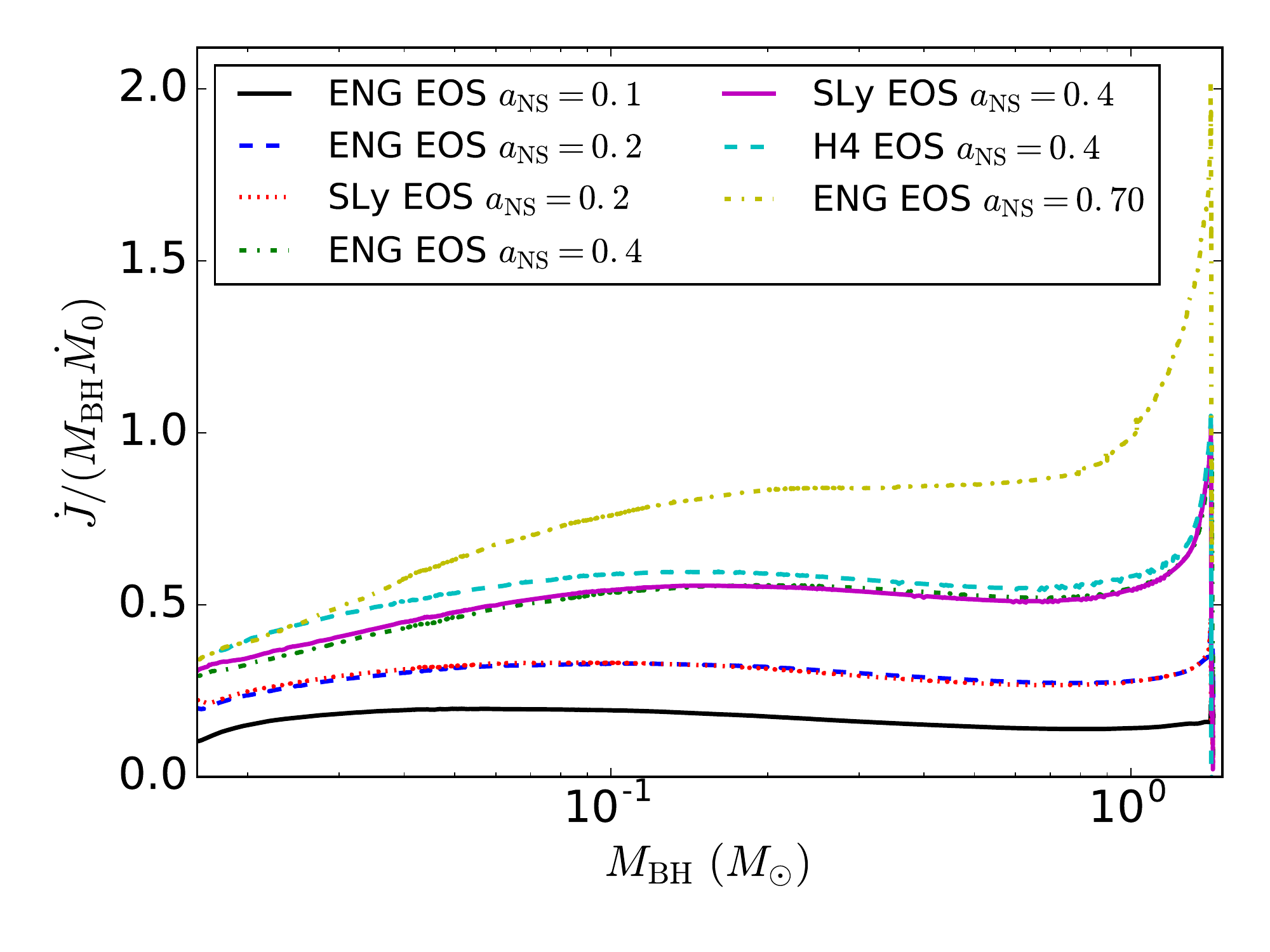}
\end{center}
\caption{
    Top: The rest-mass accretion rate for the BH, normalized by the area of the
    BH.  For comparison, the central value of $\rho_0 / c_s^3$ is $1.1\times$
    ($1.7\times$)
    higher for a nonspinning NS with the SLy EOS (H4 EOS)
    compared to the ENG EOS, which
    correlates with the difference in accretion rate.  Bottom: The ratio of the
    angular momentum to rest-mass accretion rate for the BH, normalized by the
    mass of the BH.  For both
    cases the quantities are shown versus the mass of the BH as it grows.
\label{fig:acc_rate}
}
\end{figure}

The BH accretion follows the Bondi relation $\dot{M}_0 \propto A_{\rm BH}$ to a
very good approximation all the way up to $M_{\rm BH}\sim M_{\rm NS}/2$ as
illustrated in the top panel of Fig.~\ref{fig:acc_rate}, and the differences
between the accretion rates with different EOSs fall into line with what one
expects from evaluating the central values of $\rho_0 / c_s^3$ for the
respective isolated NS solutions (also consistent with Bondi).  
Again, the matter accretion rate is largely independent of the NS spin.

More interesting is the rate at which angular momentum is accreted relative to
rest mass, as shown in the bottom panel of Fig.~\ref{fig:acc_rate}.  To a good
approximation, we find that this ratio $\dot{J}/\dot{M}_0$ is linearly
proportional to the BH radius, or equivalently, mass, throughout most of its
growth.  This is consistent with the above-noted fact that the BH spin settles
down to a roughly constant value as the BH grows [in such a regime $dJ_{\rm BH}=d(M_{\rm BH}^2) a_{\rm BH}$]. In particular, if we assume
that the BH angular momentum changes as $\delta J_{\rm BH}=\alpha M_{\rm BH} \delta M_{\rm
BH}$ for some constant $\alpha$, then $\delta a_{\rm BH}=\delta M_{\rm
BH}(\alpha-2a_{\rm BH})/M_{\rm BH}$, i.e. $a_{\rm BH}$ will have a stable
equilibrium value at $\alpha/2$.
Figure~\ref{fig:acc_rate} indicates that this proportionality constant is
relatively insensitive to the EOS. 

Examining the angular dependence of the flux through the BH horizon, we find
the rest-mass accretion to be approximately spherical, while the ratio of
the angular momentum to rest-mass accretion rate of an area element on the BH
horizon is approximately proportional to $\sin^2 \theta$ (where $\theta$ is the
polar angle), all consistent with the BH accreting spherical shells
of mass, uniformly rotating with angular velocity $\Omega$ (i.e. the
coordinate velocity in the azimuthal direction).
With this assumption, the condition that $\dot{J}/\dot{M}_0\propto R_{\rm BH}$
is equivalent to $\Omega \propto 1/R_{\rm BH}$. 
Such a relation would follow from assuming Keplerian velocity, or taking
$\Omega$ to be proportional the horizon frequency of a BH with constant
dimensionless spin, though it is inconsistent with the uniform rotation of the 
original star. See Appendix~\ref{sec:estimates} for further discussion. 

\begin{figure*}
\begin{center}
\includegraphics[width=0.66\columnwidth,draft=false]{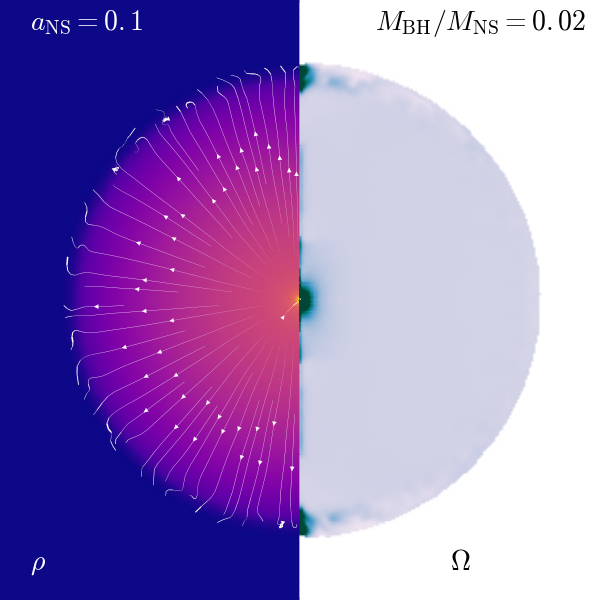}
\includegraphics[width=0.66\columnwidth,draft=false]{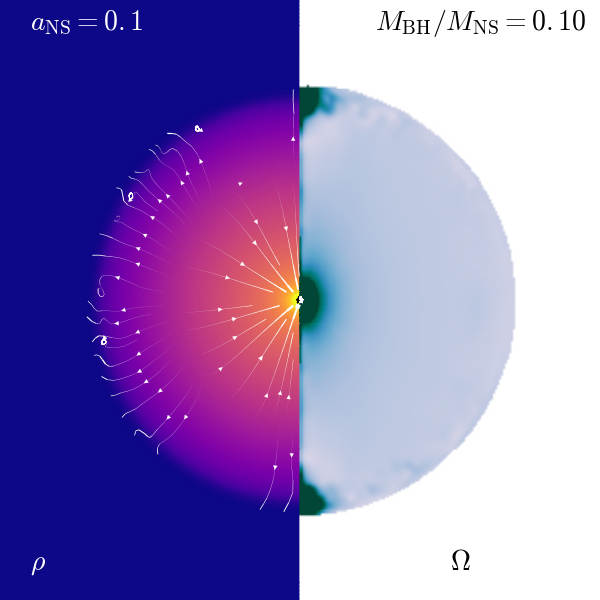}
\includegraphics[width=0.66\columnwidth,draft=false]{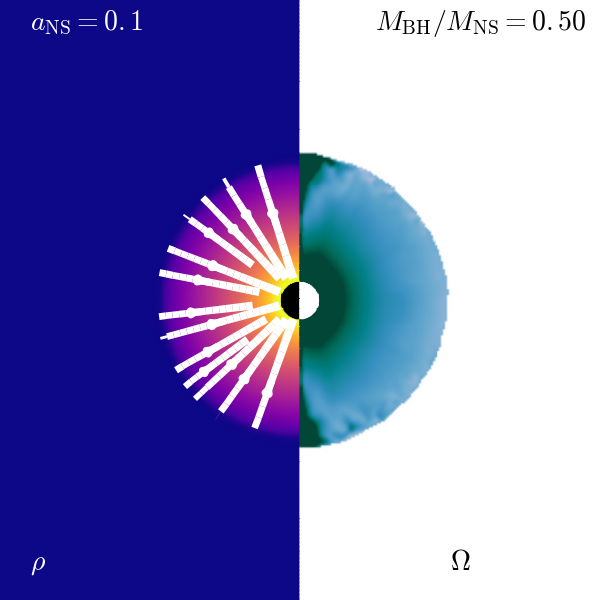}
\includegraphics[width=0.66\columnwidth,draft=false]{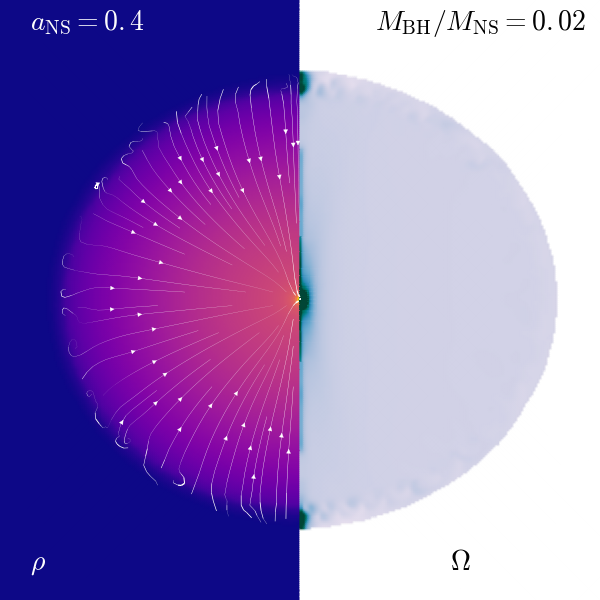}
\includegraphics[width=0.66\columnwidth,draft=false]{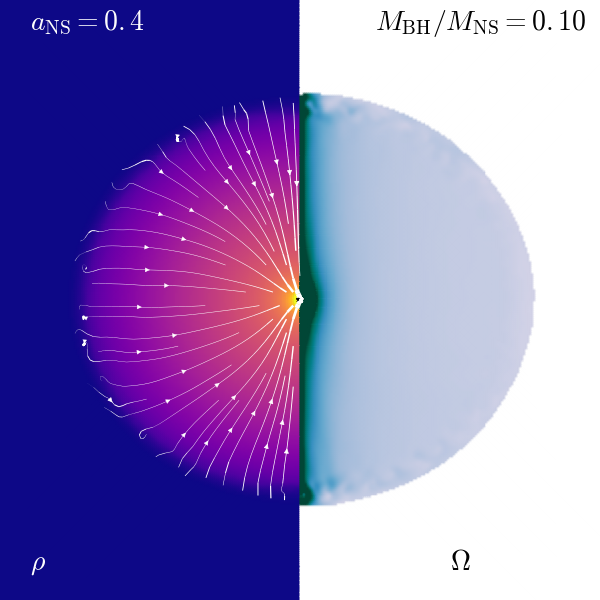}
\includegraphics[width=0.66\columnwidth,draft=false]{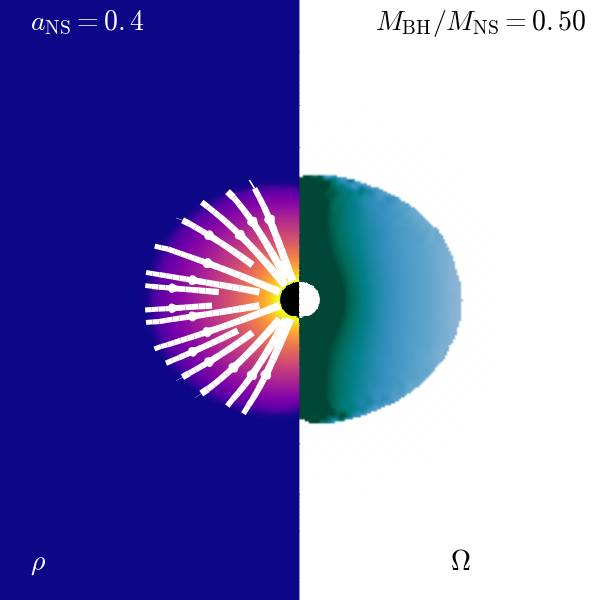}
\includegraphics[width=0.66\columnwidth,draft=false]{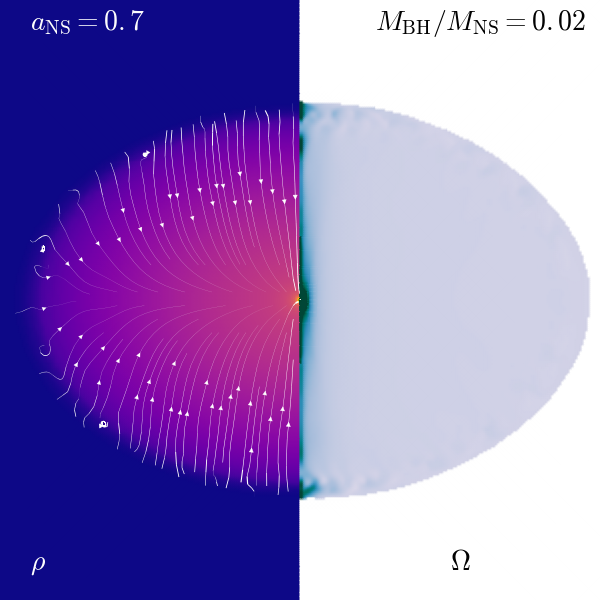}
\includegraphics[width=0.66\columnwidth,draft=false]{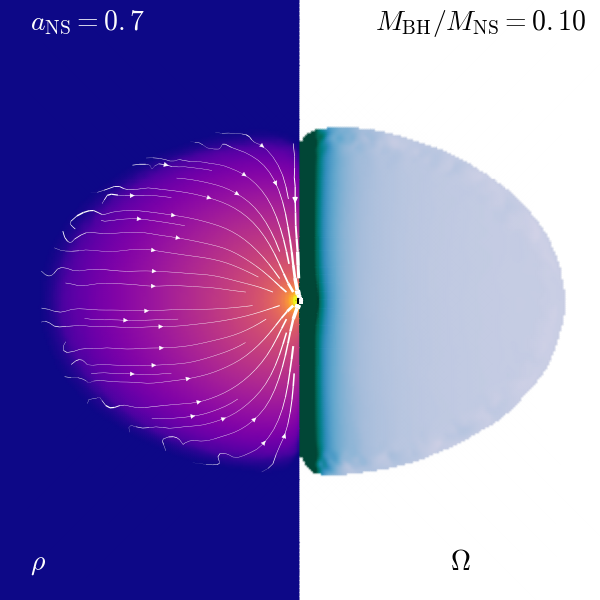}
\includegraphics[width=0.66\columnwidth,draft=false]{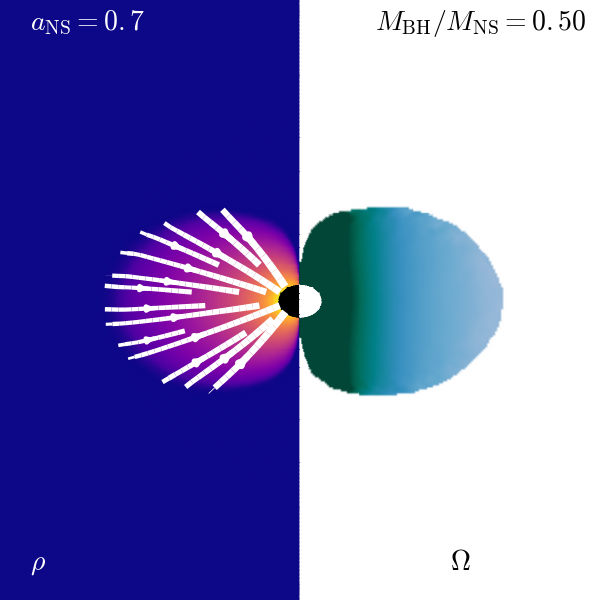}
\end{center}
\caption{
    Snapshots of density and angular velocity during the evolution of cases
    with the ENG EOS and different NS spins. The left
    side of each snapshot shows the rest-mass density (on a linear scale from 0 to $1.6\times10^{15}$ gm/cm$^3$) 
    with streamlines of the in-plane component of the velocity (with
    line width proportional to their magnitude), while the right side shows the
    angular velocity about the axis of
    symmetry. The top panels are from the case with $a_{\rm NS}=0.1$, and respectively show,
    from left to right, the points in the evolution when the BH has grown to be
    0.02, 0.1, and 0.5 times the mass of the original NS. The middle and bottom panels show the
    same for the cases with $a_{\rm NS}=0.4$ and 0.7, respectively.  The color scale for the angular 
    velocity is linear, ranging from 0 to $4\times$ the initial angular velocity of
    the NS. 
    At earlier times, the small radial velocities at the outer parts of the star
    are due to breathing modes.
\label{fig:snapshots}
}
\end{figure*}

Indeed, we find that the BH induces differential rotation in the NS, as
apparent from Fig.~\ref{fig:snapshots}, which shows snapshots
of the density and angular velocity.
There we see that the central core of the NS rotates much faster compared
to the outer part, which does not significantly increase from its original
rotation value.
Also evident from Fig.~\ref{fig:snapshots} is the high density that develops
in the vicinity of the BH. At the BH horizon we find $\rho_0\gtrsim 2\times10^{15}$ gm/cm$^3$. 
As mentioned above, these densities are 2--$3\times$ larger the initial central 
densities of the NS, and higher
even than those obtained for the maximum stable NS solutions with their respective
EOSs.
In the final phase of the NS implosion for the near breakup spin case with
$a_{\rm NS}=0.7$ (bottom-right panel of Fig.~\ref{fig:snapshots}), one
can see spherical accretion beginning to break down as the region near the rotation
axis is evacuated first, and the high angular momentum material near the equator
is the last to be accreted.
This is consistent with the late-time decrease in mass accretion rate, coupled with an increase in 
relative angular momentum accretion rate, shown in Fig.~\ref{fig:acc_rate}.

\begin{figure}
\begin{center}
\includegraphics[width=\columnwidth,draft=false]{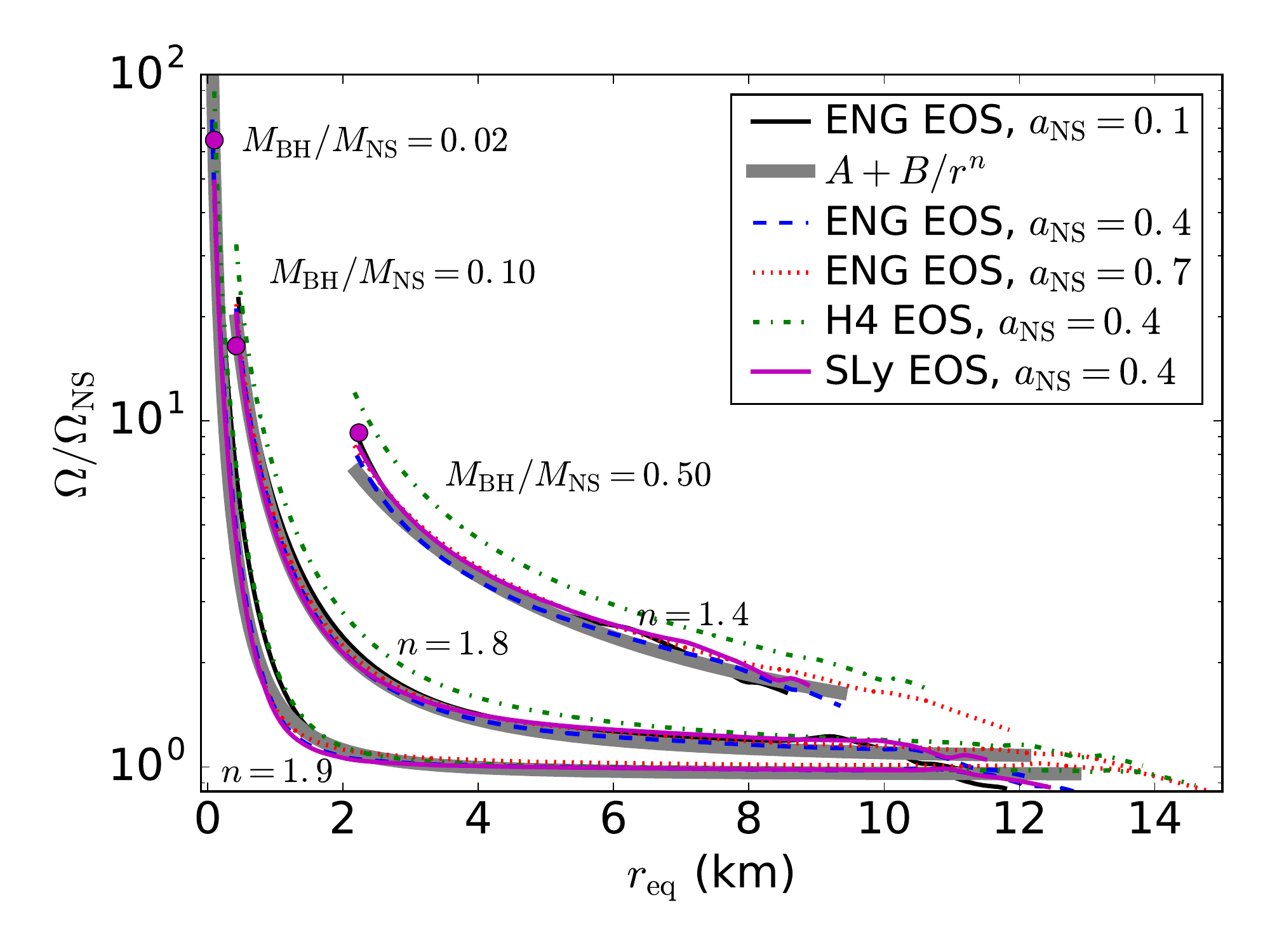}
\end{center}
\caption{
    The angular velocity on the equator as a function of proper circumferential radius.
    For each case with different EOS or NS spin, we show the rotation curve at three different
    times, corresponding, respectively, to when the BH has grown to have 0.02, 0.1 and 0.5 times the mass of initial
    NS.
    We also show fits of the form $\Omega= A+B/r_{\rm eq}^n$ to the case
    with ENG EOS and $a_{\rm NS}=0.4$ (though other cases give similar
    results). The exponent $n$ of the best-fit is found to decrease from 
    $\sim1.9$ to 1.4, as labeled in the plot.
    For the SLy EOS and $a_{\rm NS}=0.4$ case, we also include the values of the BH
    horizon frequency $\Omega_H$ for comparison (magenta dots).
\label{fig:omega_eq}
}
\end{figure}

A more quantitative illustration of the differential rotation on the
equator is shown for several cases with different EOSs and NS spins in
Fig.~\ref{fig:omega_eq}.  Here, we see that the rotation curves do not
depend strongly on the EOS, and match up well when normalized by
$\Omega_{\rm NS}$.  In Fig.~\ref{fig:omega_eq}, we also include the
best fit of the azimuthal angular velocity to the functional form
$\Omega=A+B/r^n$. At earlier times (i.e. when the BH mass is small),
this exponent $n\sim 2$, though at late times $n$
becomes smaller.  As a point of comparison, we note that the angular velocity of a
viscous fluid between two concentric spheres rotating with different velocities
has this form\footnote{Though being a Newtonian result, which ignores the effects of
gravity, and is obtained in the low Reynolds number regime, it should only be taken
as a comparison point.} with $n=3$~\cite{1987flme.book.....L}. The figure also makes
apparent the fact that the rotation rate at the accreting radius decreases, while
at the star's surface it increases. The former is expected as the angular frequency of 
the BH is $\Omega_H = a/(2 R_{BH})$---which decreases as the BH grows. Indeed, we find good
agreement with this behavior as indicated in the figure.
We also find that the rotation rate at the surface increases as the NS radius
shrinks, roughly following the 
$\Omega_s \propto R_{\rm NS}^{-2}$ behavior predicted from assuming
the angular momentum of the outer shell is constant. 
The change in $\Omega_s$ is only significant
at the last stages of the implosion, where it is accompanied by the contraction of the star,
and a strong inward radial velocity (see Fig.~\ref{fig:snapshots}). 

As a consequence of the distribution of angular momentum during the BH growth
described above, almost no matter is dynamically ejected. 
In some of the cases with higher spin, we find a few $\times10^{-5}\ M_{\odot}$ of
rest mass, flagged as unbound.
However, this is sensitive to extraction time, and seems likely to be coming from
the low density region outside the NS, rather than being ejected from the NS's surface.
Given the difficulty
in resolving the behavior in the very low density region, especially in the presence
of an artificial ``atmosphere," we cannot rule out a very small amount of unbound mass, but 
we estimate an upper bound of $M_{\rm ub} < 10^{-4}\ M_{\rm \odot}$.
(See Ref.~\cite{Camelio:2018gfc} for a discussion of these issues in a study of 
supramassive NS collapse using similar methods.)

\begin{figure}
\begin{center}
\includegraphics[width=\columnwidth,draft=false]{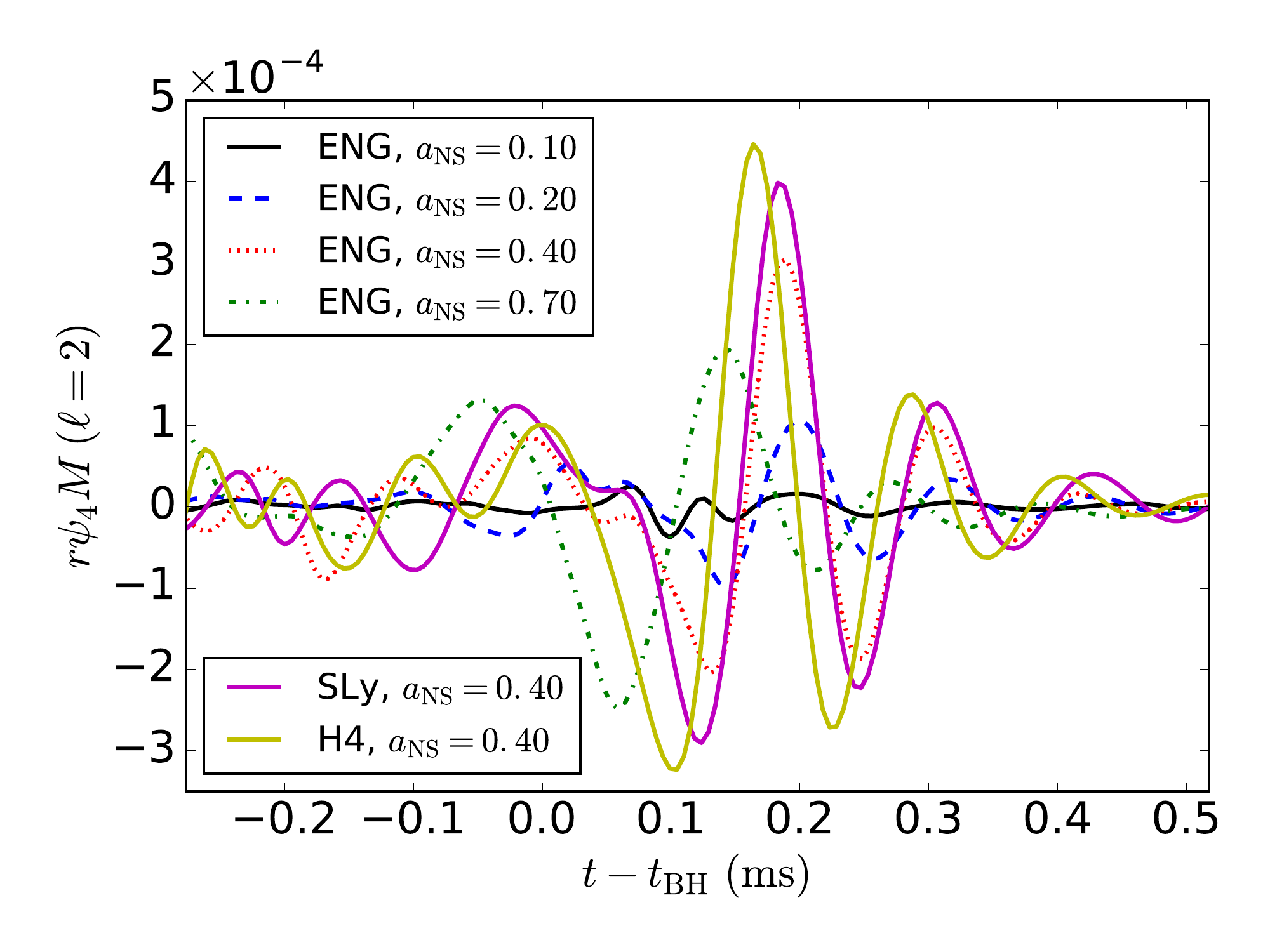}
\end{center}
\caption{
    The GW signal, in particular the $\ell=2$, $m=0$ component
    of $\psi_4$, for the collapse of different NSs. 
    The curves have been aligned in time to the look-back time where the BH 
    reaches $95\%$ of the mass of the spacetime.
\label{fig:gw}
}
\end{figure}

Finally, we examine the GW signal from this process. Though there will be no
GWs from the collapse of a NS into a BH in spherical symmetry, when there is
non-negligible NS spin there will be gravitational radiation associated 
with the quadrupole of the NS changing into that of a spinning BH.
In Fig.~\ref{fig:gw}, we can see that the GW signal exhibits 
the standard ringdown signal associated with a perturbed BH, 
with larger amplitudes for the larger spin cases.  
Unfortunately, the high characteristic frequency ($8$--$9$ kHz) of the GW signal 
makes it difficult to detect with current GW detectors. LIGO/Virgo
would likely only be able to detect such an event in the Milky Way 
or a nearby satellite galaxy (at distances of a few $\times10$ kpc).  
Though the GW amplitude is comparable to NSs induced to collapse by other
mechanisms~\cite{Duez:2005cj,Baiotti:2007np,Giacomazzo:2011cv}, the frequency
is generally higher compared to a NS collapsing because it has exceeded the 
maximum supported mass (e.g. through accretion).

\subsection*{Other electromagnetic counterparts}
Having found that the BH consumes the whole NS without giving rise to any appreciable ejecta, 
we briefly comment on possible electromagnetic transients powered by other means. 
A NS is generically magnetized with field strengths of order $B \simeq 10^{8}$ to $10^{12}$ G. 
No hair theorems imply that as the spacetime approaches that of a vacuum BH, this
magnetic field must go away.
Some of the magnetic field energy falls into the BH horizon, but
the study presented in Ref.~\cite{Lehner:2011aa} shows that a significant amount 
is radiated to infinity on dynamical timescales, giving rise to a transient signal.
The energy reservoir for such a signal is
that of the NS magnetosphere, which is of the order
\begin{equation}
	E_{M} \approx 2 \times 10^{37} \left(\frac{B}{10^{10}\ \rm G}\right)^2 \left(\frac{R}{10 \ {\rm km}}\right)^3 \mbox{erg} \ .
\end{equation}
As discussed in Ref.~\cite{Lehner:2011aa}, a fair fraction of this energy is released on timescales of
$\approx 1$ ms as a result of the collapse of the NS to a BH, resulting in a short burst with a luminosity of  
\begin{equation}
L \approx  6\times 10^{39} \left(\frac{\kappa}{0.6}\right) \left(\frac{B}{10^{10}\ \rm G}\right)^2 \left(\frac{R}{10 \ {\rm km}}\right)^3 \mbox{erg/s}
\end{equation} 
where $\kappa\approx 0.6$ is an efficiency factor.  The associated Poynting
flux follows an essentially quadrupolar distribution with most of the energy
radiated near the angles $\theta = \pm 50^o$. The environments surrounding
pulsars typically have relatively low baryon loading. The details of how this
energy is converted into specific observable signatures are certainly model
dependent (see, for instance, Refs.~\cite{Lehner:2011aa,Falcke:2013xpa} for
options ranging from short gamma-ray to radio bursts). Regardless, an
electromagnetic transient with these characteristics that lacks an associated
kilonova is arguably a clear signature of the NS implosions studied here.  As
mentioned above, it is possible that for sufficiently close systems,
multimessenger signals in both gravitational and electromagnetic bands could be
detected. Notice that NSs driven to collapse by accreting matter from a
companion or (for massive ones) angular momentum loss through spin down would
also emit in analogue ways. However, there are subtle differences that could
allow for discerning whether such signals come from the mechanism discussed
here, or these ``standard'' ones.  First, since the frequencies of GWs are
strongly correlated with the mass of the collapsing star, they would reveal
whether such a star is clearly below its maximum mass, thus favoring the
endoparasitic mechanism.  Second, such a mechanism preferentially takes
place in regions with higher dark matter density. Finally, a collapsing NS
within the standard scenario explores higher temperatures, and thus is more
likely to produce neutrinos.

\ssec{Discussion and Conclusion}
We have studied the process by which a small seed BH in the center of a rotating
NS grows, and ultimately consumes the star. In this endoparasitic process, even for NSs with
very short rotational periods, exceeding observed values, very little of the
star's material remains outside the BH ($M_{\rm ub}<10^{-4}\ M_{\rm \odot}$).  
This contrasts with the estimates of $M_{\rm ub}\sim0.1$--$0.5\ M_{\rm \odot}$
presented in Ref.~\cite{Fuller:2017uyd}. There it was
assumed that the total angular momentum would be distributed to the solid shell
of the star, which would maintain rigid rotation down to the BH horizon, plus an
outer angular shell, which could become unbound once a high enough velocity was
achieved, and that a negligible amount of angular momentum would go into the BH.
In contrast, we find that the BH obtains a non-negligible spin, and is surrounded
by differentially rotating core with much higher angular velocity than the
outer part of the NS. Furthermore, as the BH grows by a significant amount a rather 
cylindrical distribution of angular momentum is found in cases with initial higher rotational
velocity. Such behavior is consistent with the development of Ekman layers~\cite{proudman_1956}.
For the highest NS spin considered here ($a_{\rm NS}=0.7$), at late times an evacuated ``funnel'' 
region results. This behavior is expected as such a region lacks the angular momentum support to
resist prompt accretion, and is thus consumed more rapidly than the
lower latitude regions.

Here we restricted to a hydrodynamical treatment of the NS, and ignored
the effect of viscosity, magnetic fields, etc. The main justification for this
is the very short timescales in between when the BH
becomes large enough to have a significant effect on the NS as a whole, and when the NS
is completely gone. As shown here, even for EOSs with high sound speeds that give lower
accretion rates, once the BH
has consumed $1\%$ of the NS's mass, it only takes milliseconds for it to consume the rest.
The effects of sheer viscosity and magnetic braking have been studied in scenarios related to 
ours~\cite{Markovic:1994bu} in the context of a potential BH 
in the Sun, and adapted for the NS case in Refs.~\cite{Kouvaris:2010jy,Fuller:2017uyd}. 
Sheer viscosity is deduced to affect the angular rotation while $M_{\rm BH} \lesssim 10^{-3}\ M_{\odot}$, with
magnetic braking becoming the relevant effect afterwards. These works conclude that
the viscosity and magnetic braking ensure that rotation does not halt the infall of matter into the BH, and both Bondi accretion 
and uniform rotation are preserved through essentially the full lifetime of the consumption process. 
Notice Newtonian estimates miss, in particular, dragging by the BH's rotation and the accretion process
itself. Furthermore, during the rapid
evolution ensuing after the BH mass satisfies $M_{\rm BH} \gtrsim 10^{-2}\ M_{\odot}$, the 
\Alfven timescale $\tau_B \simeq (10^{12}\ {\rm G}/B)$ s becomes too long for magnetic braking
to operate efficiently\footnote{Radiation effects, e.g., from neutrino emission, would operate on far longer timescales
and can be safely ignored.}. Our hydrodynamical calculations, which begin when the BH mass is already a hundredth of the NS mass
with uniform rotation for the NS, are thus a realistic representation of the state the preceding dynamics---beginning
with a BH seed with mass $\lesssim10^{-8} M_{\odot}$---should give rise to.

We carried out a thorough study of different initial values
for the BH mass and spin (see Appendix~\ref{sec:ics}), which we find all give
similar results. Moreover, we argue that our results should extrapolate to
smaller BH masses (at least until other processes which we neglect here, e.g. 
magnetic braking or viscosity, become important) based on the simple power law scaling relationships we
observe for the rates of accretion of mass and angular momentum with the BH
mass (approximately $\dot{M}_0\propto M_{\rm BH}^2$ and $\dot{J}\propto M_{\rm
BH}^3$), which we find hold until the very final stages of the NS implosion,
and which give nearly constant dimensionless BH spin. 

Our results put a damper on the ``quiet kilonovae" scenario for observing NS
implosions---i.e. a kilonovae that is not accompanied by the significant GW
signal of a BH-NS or NS-NS merger---as well as for placing bounds on their
rates through the abundance of r-process material. However, several
observational possibilities still remain. We predict that this mechanism should
produce a population of BHs whose mass and angular momentum distribution should
tightly match that of the NSs which gave rise to them.  If these BHs were to
subsequently merge with a compact object companion, as discussed in the
introduction, their BH nature could in principle be deduced from the GW signal
and/or electromagnetic counterpart (or lack thereof).  In principle, the GW
signal from the NS implosion itself would encode the mass and spin of the newly
formed BH (and distinguish from NSs collapsing due to becoming too massive),
and could be observed with current detectors within our own Galaxy or nearby.
Future GW detectors, especially those targeting the kilohertz frequency
band~\cite{Martynov:2019gvu}, could improve this. 

As also discussed, a magnetized NS 
has to shed its magnetosphere upon conversion to a BH, which leads to an
electromagnetic outburst on millisecond timescales, though with unknown frequency. 
An interesting avenue for future work would be to include magnetic fields 
in a study similar to this one, in order to track the possible enhancement due to the differential rotation
of the NS, and better model the release of the electromagnetic energy as the BH grows large.
Energetically, such events are plausible sources for
FRBs, and even short gamma-ray bursts, which would be unaccompanied by a kilonova. New observatories like
CHIME~\cite{Amiri:2019qbv}, HIRAX~\cite{Newburgh:2016mwi},
SKA~\cite{Carilli:2004nx}, FAST~\cite{Nan:2011um}, and others are rapidly
increasing the number of observed FRBs, and providing crucial clues to their
source(s).  Upcoming radio surveys will also take a more accurate census of the
pulsar population in the galactic core, which will place bounds on, or provide
evidence for, the scenario studied here. Likewise, gamma-ray
detectors like Fermi, VERITAS, MAGIC, and 
HAWC~\cite{2017ApJ...848L..14G,HAWC:2018eaa,Alfaro:2018xep,Archer:2019mxj} could potentially
help identify these events.

Finally, we mention that it is intriguing that these systems probe higher
densities than even massive NSs, about an order of magnitude above nuclear
density, with this high density persisting during the accretion phase. Though
it is unclear whether there will be any special observational signature (e.g.,
through neutrino seismology), since the material is in the process of falling
into the BH, the behavior of matter at these densities is highly unknown, and
may not be otherwise probed.

\ssec{Acknowledgements}%
We thank Joe Bramante for helpful discussions.
W.E. and L.L. acknowledge support from an NSERC Discovery grant. 
L.L. acknowledges CIFAR for support. This research was
supported in part by Perimeter Institute for Theoretical Physics.  Research at
Perimeter Institute is supported by the Government of Canada through the
Department of Innovation, Science and Economic Development Canada and by the
Province of Ontario through the Ministry of Research, Innovation and Science.
This research was enabled in part by support provided by SciNet
(www.scinethpc.ca/) and Compute Canada (www.computecanada.ca).

\appendix


\section{Details of initial conditions}
\label{sec:ics}
In this section, we give details on the initial data we use to describe a
rotating NS with an interior BH.  We construct solutions to the Einstein
constraint equations by solving the conformal thin-sandwich equations as
described in Ref.~\cite{idsolve_paper}. We choose the free data for these
solutions by combining a uniformly rotating NS solution with mass $M_{\rm NS}$
and dimensionless spin $a_{\rm NS}$, obtained using the RNS
code~\cite{Stergioulas:1994ea}, with a rotating BH solution with mass $M_{\rm
BH}$ and dimensionless spin $a_{\rm BH}$, in Kerr-Schild coordinates, 
as we describe below.

We begin with a NS solution in spherical-polar type coordinates with the radial
coordinate $R$ chosen such that $2\pi R \sin \theta$ is the proper circumference of any point
rotated around the axis of symmetry.
We choose a radius $R_{m}$ where we match on to the BH solution,
and rescale the time coordinate of the NS solution by a constant factor
so that is agrees with the value for the lapse for a Boyer-Lindquist BH on the equator
at $R_{m}$. Then we apply the coordinate transformation that goes from Boyer-Lindquist
to Cartesian Kerr-Schild coordinates for a BH with parameters $M_{\rm BH}$ and $a_{\rm BH}$ to the
NS solution to obtain the metric $g_{\rm ab}^{NS}$. 
We combine this metric with the BH metric in Kerr-Schild coordinates using a transition function:
$g_{ab}=f(R) g^{\rm NS}_{ab}+\left[1-f(R)\right] g^{\rm BH}_{ab}$ (since in the absence of spherical symmetry, the two solutions
do not agree at $R=R_m$) with $f(R) = T\left((R-R_m+\delta R)/(2 \delta R)\right)$
and where 
\begin{equation}
	T(x)=
\left\{
	\begin{array}{lll}
		0  & \mbox{if } x<0\\
		x^3(6x^2-15x+10) & \mbox{if } 0\leq x \leq 1 \\
		1 & \mbox{if } x > 1 
	\end{array} 
\right. 
\end{equation}
interpolates between 0 and unity with continuous first and second derivatives.
Finally, we rescale the time coordinate of the combined solution by an overall factor
so that the lapse goes to unity at spatial infinity.
From this solution we calculate the free data metric functions.
We want $R_{m}$ to be intermediate in scale between the BH and
the NS radius, and $\delta R$ to be intermediate in scale between
the BH radius and $R_{m}$, though the resulting dynamics are
not very sensitive to the exact values.
We use $R_{m}\approx 40M_{\rm BH}$ and $\delta R = 0.1 R_{m}$.

Similarly for the matter, we combine the density and velocity profile of the
rotating NS solution with a quasiequilibrium test fluid solution on the BH
background.  For the latter, we assume constant $h u_t$ (where
$h=1+\epsilon+P/\rho_0$ is the specific enthalpy), where the constant value
comes from the NS solution at $R=R_{\rm m}$ on the equator.
We also assume the same constant angular velocity throughout. 
Very near the BH ($R<3 M_{\rm BH}$), we set the density to 0.
Again, we combine these
two solutions using the same transition function, e.g. $\rho_0 = f(R)\rho_0^{\rm NS}+\left [1-f(R) \right ]\rho_0^{\rm BH}$.
From this, we calculate the conformal energy and momentum density, which are
used in solving the conformal thin-sandwich equations.

\begin{figure}
\begin{center}
\includegraphics[width=\columnwidth,draft=false]{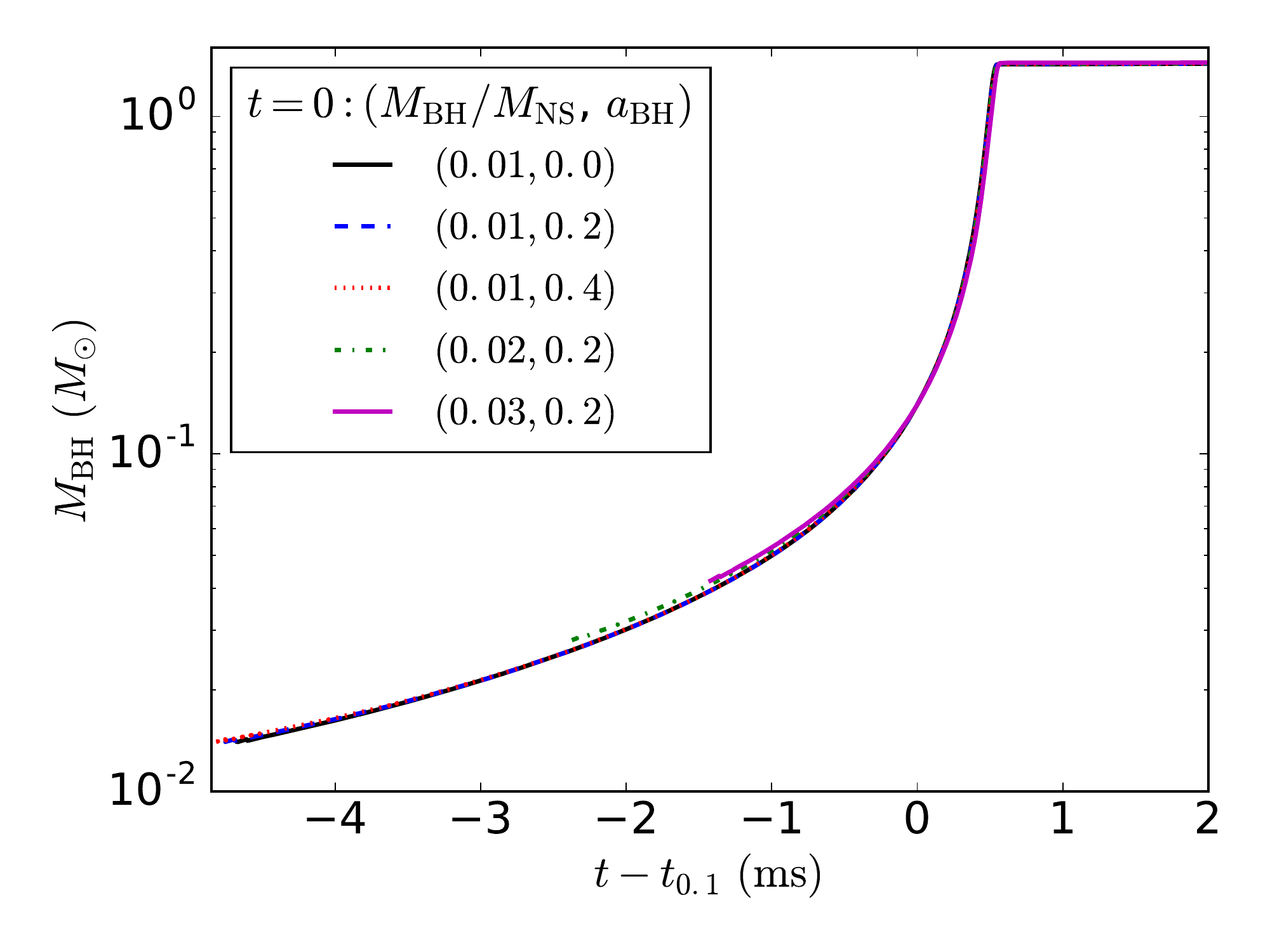}
\includegraphics[width=\columnwidth,draft=false]{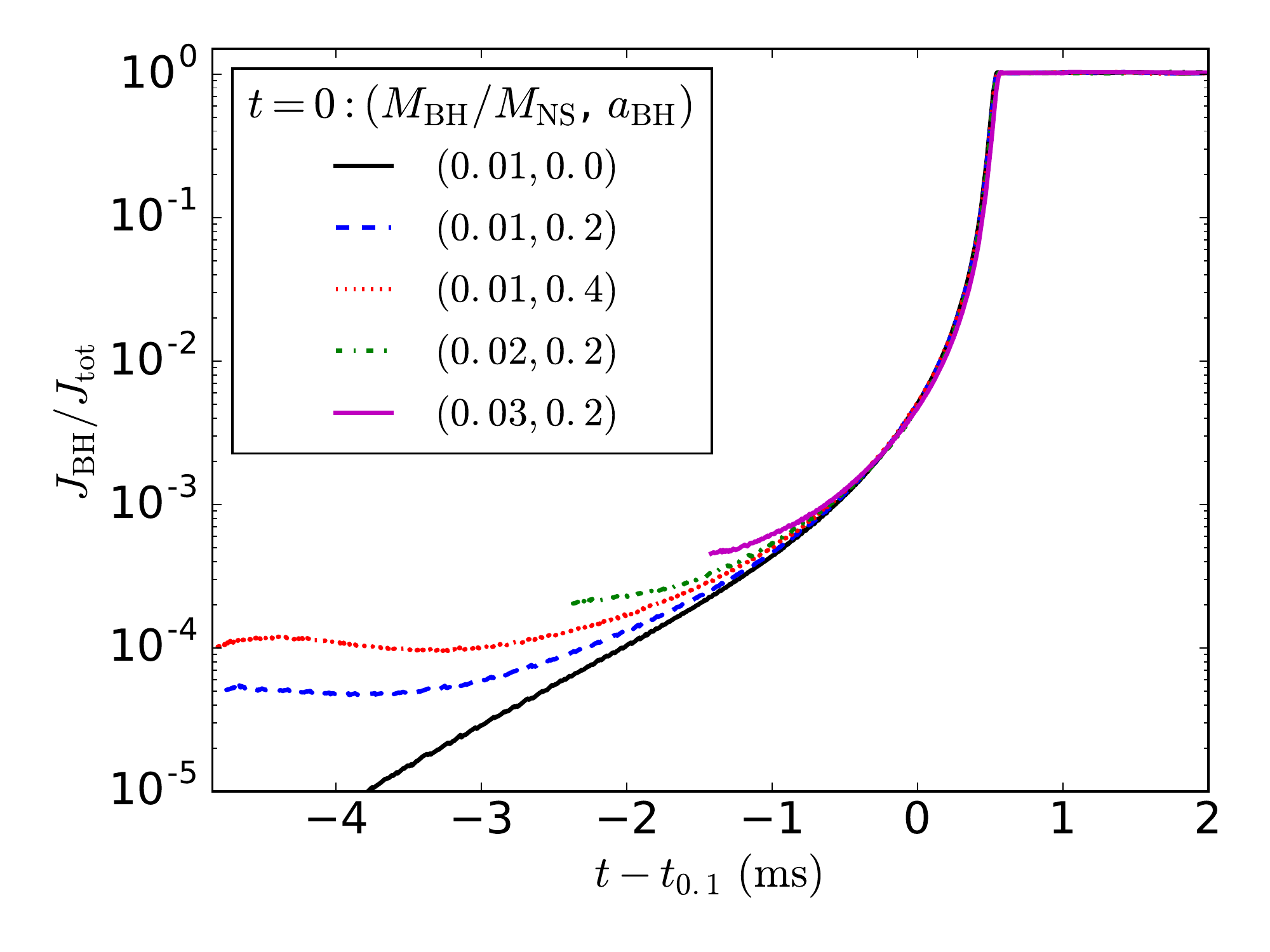}
\end{center}
\caption{
    The mass (top) and angular momentum (bottom) of the BH as a function of time for cases with 
    the ENG EOS and $a_{\rm NS}=0.4$, but different initial masses and spins for the BH.
    The curves have been aligned at the time where $M_{\rm BH}=0.1M_{\rm NS}$.
\label{fig:mbh_mbhi}
}
\end{figure}

The corrections we find from solving the constraints are small; e.g.
the maximum difference from unity of the conformal factor over the domain is
$\max |\Psi-1|\sim 0.01$.
There is an initial transient when evolving these solutions since, for
example, we do not include an inward radial velocity. However, we find the
solution settles down to steady accretion in a few sound crossing times. 
We verify that our initial data are good enough, and that we start sufficiently
early in the growth of the BH, by comparing solutions
with different initial values of $M_{\rm BH}$, ranging from 0.01 to 0.03
$M_{\rm NS}$. As shown in Fig.~\ref{fig:mbh_mbhi}, the subsequent growth 
is very similar in all cases.

Another issue is what value should be chosen for $a_{\rm BH}$. For simplicity,
we just set $a_{\rm BH}=a_{\rm NS}$. With this value, $a_{\rm BH}$ decreases
during the BH's initial growth, but as evident in the bottom panel of
Fig.~\ref{fig:mbh_mbhi}, the initial value of $a_{\rm BH}$ becomes unimportant by
the time the BH doubles or triples in mass.

\section{Numerical convergence and comparison of axisymmetry to 3D}
\label{sec:conv}
In axisymmetry, our numerical grid spans the half-plane $(x,z)\in[0,\infty)\times(-\infty,\infty)$.
For our default resolution we use 13 levels of 2:1 mesh refinement with $257\times129$
points on the coarsest level, and a resolution of $dx\approx 3\times10^{-4} M_{\rm NS}$
on the finest level.
The mesh refinement levels are dynamically adjusted according to truncation error
estimates in the metric functions. The initial threshold is set so that the BH region
is completely covered by the finest level during the first part of the evolution, and
is only dropped after the BH doubles in size.

\begin{figure}
\begin{center}
\includegraphics[width=\columnwidth,draft=false]{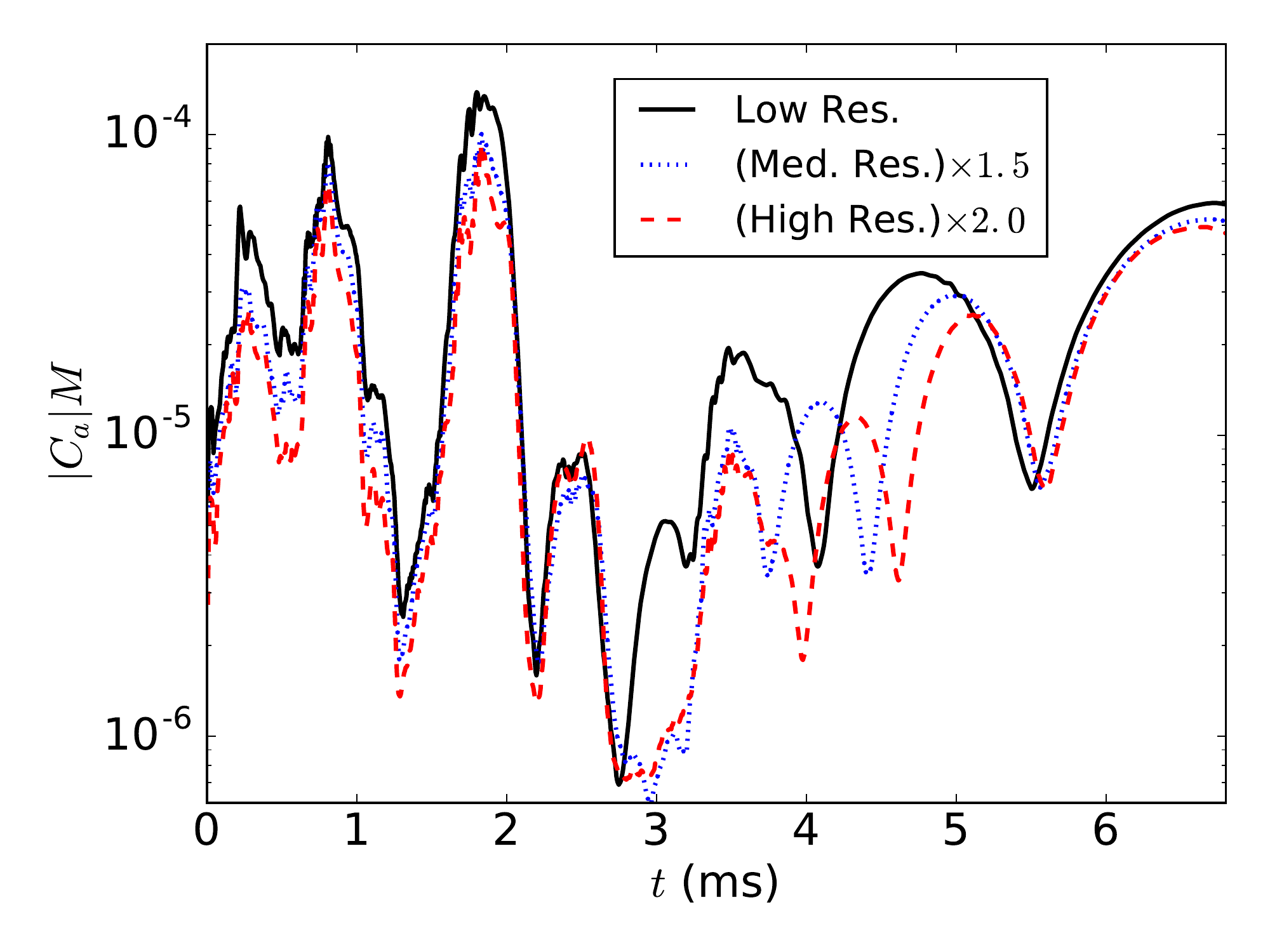}
\end{center}
\caption{
The convergence of the L2 norm of the generalized
harmonic constraint violation $C_a=H_a-\Box x_a$ 
(average value in the $[-15M_{\rm NS},15M_{\rm NS}]\times[0,15M_{\rm NS}]$ central portion of the domain) 
for the case with the ENG EOS and $a_{\rm NS}=0.4$.
The results have been scaled assuming first order convergence, though at early times
the convergence in the quantity is close to second order.
\label{fig:cnst}
}
\end{figure}

\begin{figure}
\begin{center}
\includegraphics[width=\columnwidth,draft=false]{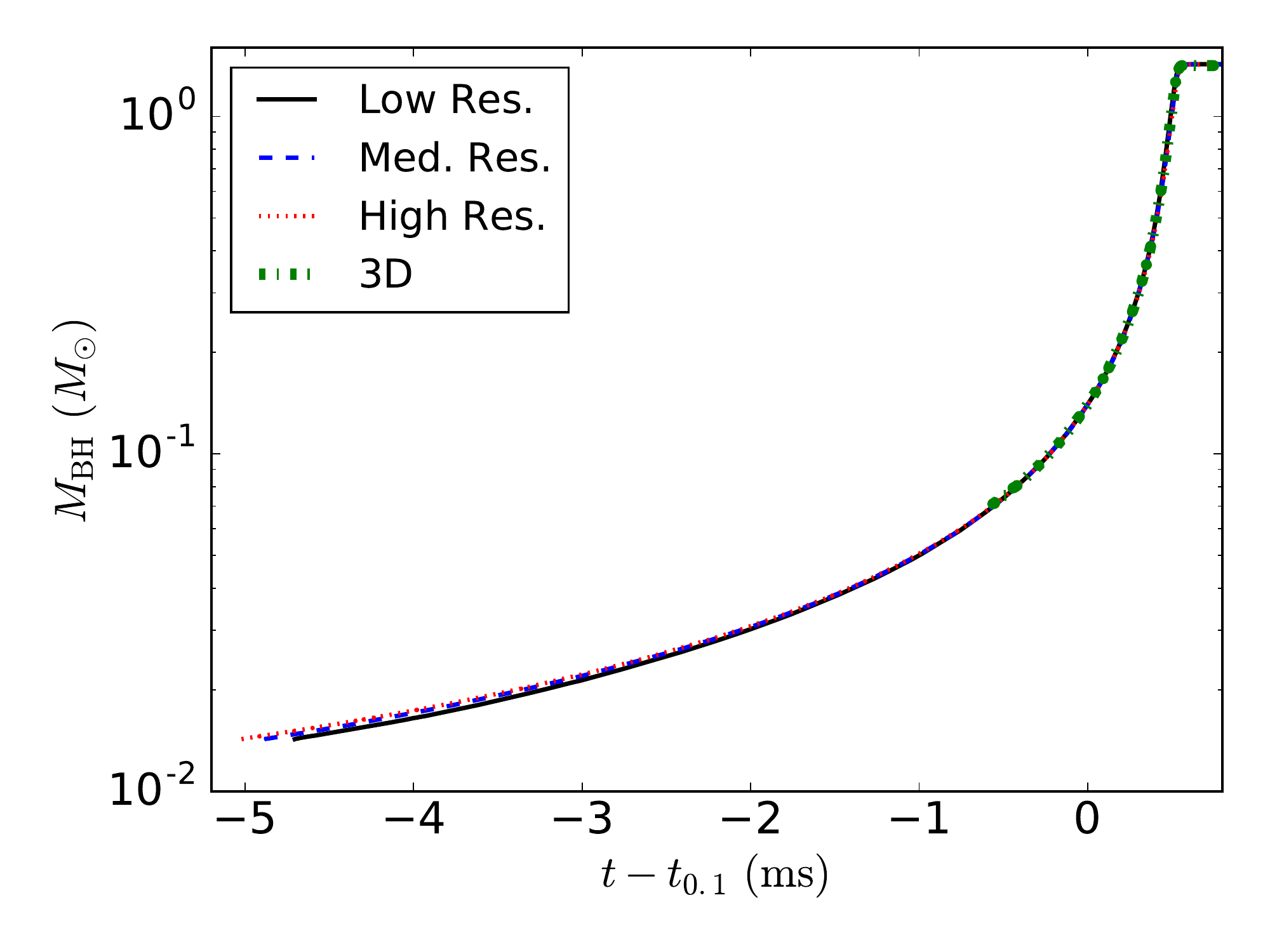}
\includegraphics[width=\columnwidth,draft=false]{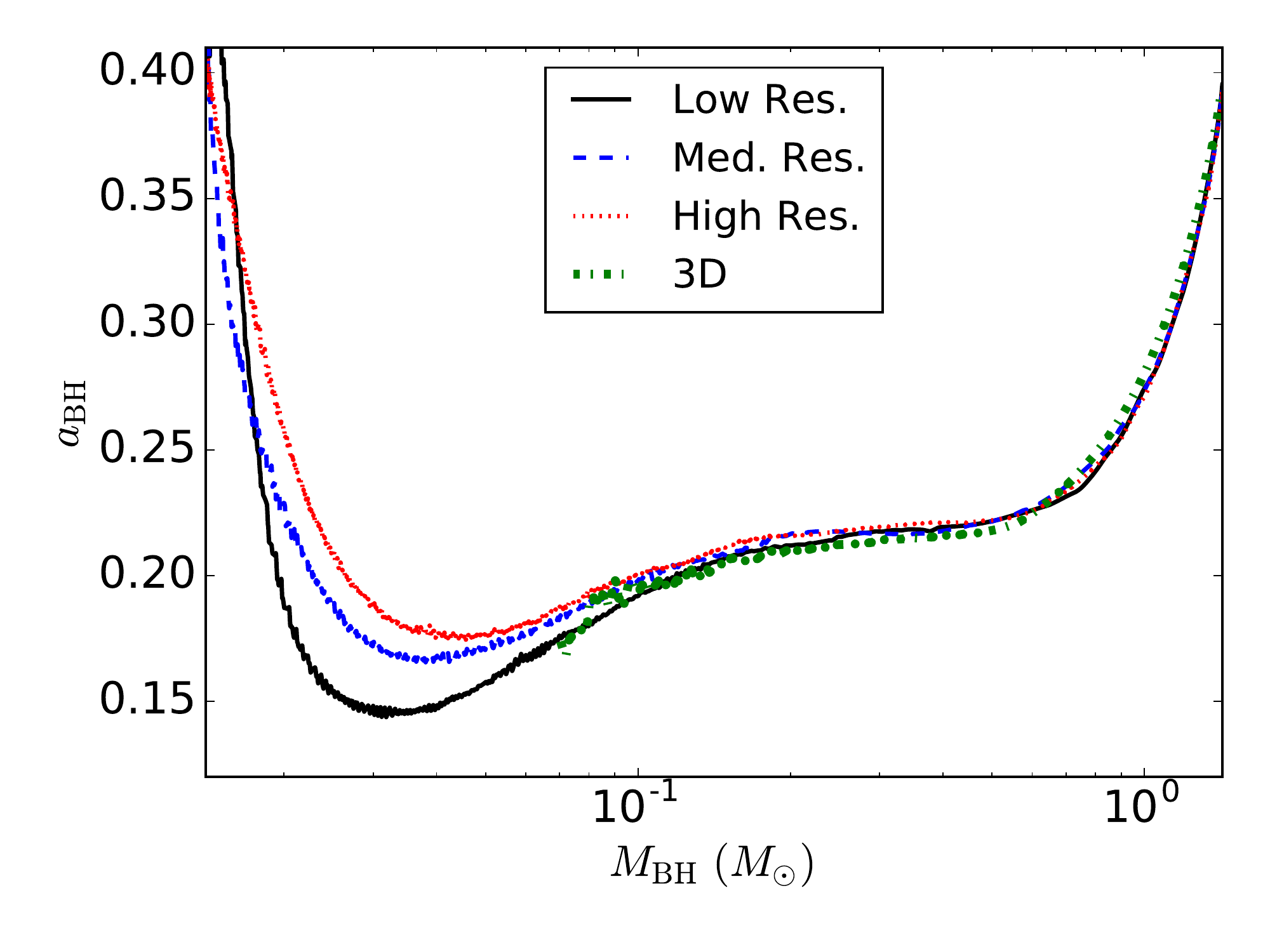}
\end{center}
\caption{
    Top:
    The mass of the BH as a function of time for the case with 
    the ENG EOS and $a_{\rm NS}=0.4$ for several different numerical resolutions. 
    We also include the curve from a three-dimensional (i.e., without assuming axisymmetry) 
    evolution of the final $\sim$ ms of the collapse that uses the axisymmetry
    evolution for initial data.
    The curves have been aligned at the time $t_{\rm 0.1}$ where $M_{\rm BH}=0.1M_{\rm NS}$.
    The primary effect of finite resolution is to decrease $t_{\rm 0.1}$, with the
    Richardson extrapolation (consistent with first order convergence) using all three resolutions giving $t_{\rm 0.1}\approx 5.4$ ms.
    Bottom: The dimensionless spin of the BH versus its mass for the same cases.
\label{fig:mbh_res}
}
\end{figure}

To establish convergence, and to estimate truncation error, we also perform
simulations of the ENG EOS, $a_{\rm NS}=0.4$ case (with $M_{\rm BH}=0.01M_{\rm NS}$
and $a_{\rm BH}=a_{\rm NS}$) at 1.5 and $2\times$ higher resolution (i.e., 
with grid spacing that is $2/3$ and $1/2\times$ smaller).
In Fig.~\ref{fig:cnst}, we demonstrate the convergence of the constraints.
We show the evolution of the BH mass and spin for this resolution study
in Fig.~\ref{fig:mbh_res}. The main effect of finite resolution is a slight
overestimate of the BH accretion rate, and an underestimate of the BH 
spin.

We also simulate this same case in full 3D, i.e. without explicitly enforcing
axisymmetry, in order to check whether there are any nonaxisymmetric instabilities.
We use resolution equivalent to the default resolution used in axisymmetry.
Because of the expense of evolving in full 3D, we begin the evolution when $M_{\rm BH}=0.05M_{\rm NS}$,
using the axisymmetric evolution to determine the initial conditions.
As also shown in Fig.~\ref{fig:mbh_res}, we find almost no difference in the growth of the BH.

\begin{figure}
\begin{center}
\includegraphics[width=\columnwidth,draft=false]{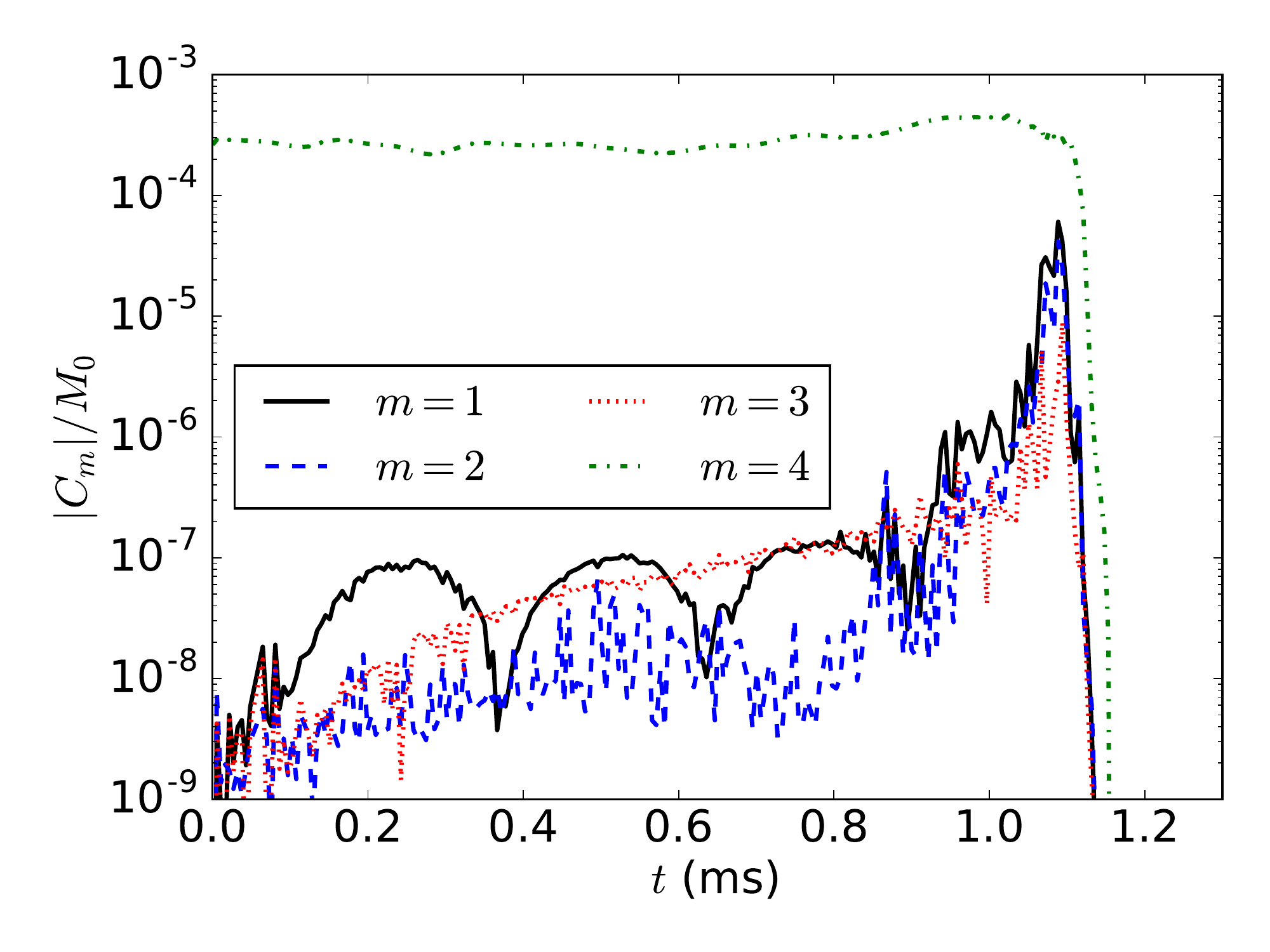}
\end{center}
\caption{
    The nonaxisymmetric density modes, normalized by the total initial rest mass of the star, 
    in the 3D simulation as a function of time.
    The sudden decrease at $t\approx 1.1$ ms corresponds to time when the NS is almost
    completely consumed by the BH.
\label{fig:nonaxi}
}
\end{figure}

We also check for nonaxisymmetric modes by computing the azimuthal decomposition of the density:
\begin{equation}
	C_{\rm m} = \int \rho_0 u^t \sqrt{-g}e^{i m \phi} d^3x
\end{equation}
where $g$ is metric determinant, 
as was done in Refs.~\cite{Paschalidis:2015mla,East:2015vix} in order to study
the one-arm mode instability in hypermassive NSs.  We plot $C_m$, for $m=1$ to 4,
as a function of time in Fig.~\ref{fig:nonaxi}.  The largest component is the
$m=4$, which is expected from the truncation error associated with discretizing
a sphere on a Cartesian grid, and all the modes remain negligible throughout.
There is no evidence for any growing nonaxisymmetric modes before the time
that the BH has consumed order unity of the NS's mass.

\section{Estimates}
\label{sec:estimates}
To gain insight into the accretion process and the consequences for the
dynamical behavior, it is convenient to visualize the NS-BH interaction in
``shellular'' terms. In particular, we further distinguish three distinct
regions: (i) a BH with mass and spin parameters $M_{\rm BH}$, and $a_{\rm BH}$;
(ii) an about to be accreted thin shell at radius $R_s \approx R_{\rm BH}$ with mass $\delta M_s$ and
angular momentum $\delta J_s$ that is uniformly rotating with angular
rotational velocity $\Omega_s$; and (iii) the rest of
the star, with mass $M_R$, and angular velocity $\Omega_R(r)$.  At the onset,
the angular velocity is constant and set by the NS spin:
$\Omega_s=\Omega_R(r)=\Omega_{\rm NS}$.

The shell being accreted by the BH induces
a change of mass and angular momentum. The amount of the latter is constrained by
the BH satisfying the Kerr bound $a_{\rm BH} \le 1$ (we assume throughout this section 
that cosmic censorship holds).
Let us analyze the change in the dimensionless BH spin due to accretion. From the definition of $a\bh\equiv J\bh/M\bh^2$, 
\begin{equation}
\delta a\bh = M\bh^{-2} \left(\delta J\bh - 2 a\bh M\bh \delta M\bh \right ) \ . \label{delta1}
\end{equation}
Approximating the angular momentum of the shell using the Newtonian expression for a
rotating spherical shell $\delta J_s= (2/3) \delta M_s R\bh^2 \Omega_s$, upon accretion 
of the whole shell the change in dimensionless BH spin is:
\begin{equation}
\delta a\bh = \frac{\delta M_s}{M\bh} \left( \frac{2}{3} \frac{R\bh^2 \Omega_s}{M\bh} - 2 a\bh \right ) \ . \label{delta2}
\end{equation}
Since $R\bh$ grows with $M\bh$, for $\Omega_s \propto M\bh^{p}$, it is clear that if $p \ge -1$, as the BH 
grows its spin will remain below the Kerr bound.
With this observation, one can derive also what is expected for $\delta J/(\delta M R\bh)$ for the BH.
Continuing with our Newtonian approximation, we can estimate the accretion of mass and angular momentum
as a function of polar angle of a shell of width $\delta R$,
\begin{eqnarray}
    \frac{d \delta M}{d \theta} &=& R^2 \delta R 2\pi \sin\theta \rho \ , \\
    \frac{d \delta J}{d \theta} &=& (R^2 \delta R 2\pi \sin\theta \rho) R^2 \sin^2\theta \Omega_s \ .
\end{eqnarray}
At constant $R$, these expressions imply $(d\delta J/d\theta)/(d\delta M /d\theta)\propto \sin^2\theta$ as observed in our simulations. If at
any point the angular rotation $\Omega_s$ were to approach breakup, 
the assumption that the BH is accreting rigidly rotating spherical shells should
break down, and the accretion of material near the equator would be strongly affected.

Several possible values for $p$ in the relation $\Omega_s \propto M\bh^p$ are as follows: 
\begin{enumerate}[(i)]
\item $p=0$: The angular velocity of the accreted material remains constant at
    the initial value $\Omega_{\rm NS}$ set by NS spin. After
        the initial transient stage, this is not consistent with what
        is seen in the simulations. 
\item $p=-1$: This would arise if the angular rotation at the accretion radius
    obeys a Keplerian relation $\Omega_K=\sqrt{M\bh/R\bh^3}$, or is given by the
        BH horizon rotational rate $\Omega_H = a\bh/(2 R\bh)$. This behavior is
        consistent with observations in the intermediate regime where the spin
        remains nearly constant.
\item $p=-4/3$: This would arise if every spherical shell 
        fell from an initial radius $R_i\sim[3M\bh/(4\pi\rho)]^{1/3} \propto M\bh^{1/3}$
        (assuming constant density) with its angular rotational velocity
        growing like $\Omega_s=\Omega_{\rm NS}(R_i/R\bh)^2$ to maintain constant angular momentum.
        As noted above, such a power would mean that $a\bh$ would grow towards unity when the BH was sufficiently small. 
	It would also mean that $\delta J/(\delta M R\bh)$ would decrease as the BH grows, which is not
	observed in the simulations. 
\end{enumerate}

We are thus left with the following picture. At early
stages, the BH grows in mass and its spin parameter either decreases or increases as dictated
by Eq.~(\ref{delta2}) depending on its initial value. Then, an intermediate 
stage takes place where the BH has roughly constant dimensionless spin, and 
grows keeping $\delta J / (\delta M R\bh)$ roughly constant. The rest of the star
increases its angular rotation rate as the star is gradually consumed by the BH. Our simulations show
$\Omega\bh$ decreases during this stage as $R\bh^{-1}$ while $\Omega(r=R_s)$ 
(the rotational velocity of the star's surface) increases. The dynamics induce a radial dependence 
on $\Omega(r)$, which we fit to an  expression $\Omega = A + B/r^{n}$.

\bibliographystyle{apsrev4-1.bst}
\bibliography{ref}

\end{document}